\documentstyle[manuscript,aps,epsf]{revtex}

\begin{document}

\title{CONVECTIVE MOTION \\ IN A VIBRATED GRANULAR LAYER }
\author{
A. Garcimart\'{\i}n \thanks{electronic mail: angel@fisica.unav.es}, 
D. Maza, 
J.L. Ilquimiche \thanks{Permanent address: Laboratorio de Optica 
F\'{\i}sica y L\'aseres,  Facultad de Ciencias F\'{\i}sicas 
y Matem\'aticas, Universidad Nacional de La Libertad, 
Avda. Juan Pablo II s/n, Trujillo, Per\'u} 
and I. Zuriguel
\\
{\it Departamento de F\'{\i}sica, Facultad de Ciencias, 
Universidad de Navarra, 
\newline E-31080 Pamplona, Spain.}
}

\date{\today}

\maketitle

\vspace{-1cm}
\begin{abstract}

\noindent
Experimental results are presented for a vertically
shaken granular layer.  In the range of accelerations
explored, the layer develops a convective motion in 
the form of one or more rolls. The velocity 
of the grains near the wall has been measured. It grows
linearly with the acceleration, then the growing rate 
slows down.
A rescaling with the amplitude of the wall velocity and
the height of the granular layer makes all data collapse
in a single curve. This can provide insights
on the mechanism driving the motion.
45.70.-n Granular systems

\end{abstract}

\newpage

Granular materials display different behaviors depending 
on the circumstances: they can be 
assimilated to solids, liquids or gases \cite{Jaeger}.
Yet the features and the origin of their behavior are 
far from understood and are the subject of current 
research. The focus of this work is on the motion that 
arises in a layer of noncohesive granular material when 
submitted to a vertical vibration. Such layers are known 
to develop large scale collective motions 
\cite{Jaeger,Duran,Nagel}. Once a certain threshold in 
the acceleration is reached (close to $\mathbf{g}$, the acceleration
of gravity) grains begin to move orderly. This movement
takes a form in many respects similar to that obtained
in convecting fluids heated from below (the Rayleigh-B\'enard 
instability \cite{Cross}). Other instabilities, such as 
parametric wave patterns and the formation of a heap, have 
also been reported for granular layers \cite{Melo,Laroche}. 
Here we are dealing only with the closed flow of granular 
material, at velocities much smaller than the maximum speed 
of the vibrating container.

This convective flow is usually downward near the wall, and 
the grains rise at the center of the container \cite{Nagel}. 
For a given frequency, if the acceleration is increased 
there is a bifurcation towards a pattern of rolls: the 
velocity oscillates spatially along one preferred direction 
\cite{Aoki2}. This symmetry-breaking bifurcation is known 
to display strong hysteresis. Further augment of the 
acceleration leads to a disordered state.

Some mechanisms have been proposed to explain this 
phenomenon. The most obvious is the friction of the 
grains with the walls, as suggested by Knight et al. 
\cite{Nagel}. Other effect concerns the Reynolds dilatancy 
notion: a tightly packed granular material must increase its 
volume if it has to change shape. During each cycle, the 
material undergoes a process involving a compaction and a 
dilatation that has been proposed to be at the origin of 
the motion \cite{Rajchenbach1}. And yet another analysis
---based on the assumption that a ``granular temperature''
drives the flow, as in thermal convection--- has been 
recently proposed \cite{Ramirez}. The role of the
surrounding gas has also been described in the case of
the formation of a heap \cite{Behringer}: it acts as
a lubricant; but it is generally admitted that it is not
at the origin of the motion. Notice that all these 
explanations are not mutually exclusive. While they
may all contribute to the convection, it is difficult to
ascertain their relative significance in a particular
situation.

A measurement of the velocity can clearly help to 
understand the underlying physics. But carrying out such
a measurement in a non invasive way is complicated.
The use of trace particles is possible but painful and
time-consuming. Magnetic resonance imaging has been
employed to obtain velocity profiles along a cross section
of the container \cite{Nagel}, but the setup is entangled
by the fact that an electromagnetic shaker cannot be
placed inside the machine. We have arranged an experiment
that allows us to obtain the local velocities at 
the wall of the container. While the nature of the flow
is three dimensional, the velocity at the boundaries 
provides a good way to define it.

The experimental setup consists of a cylindrical box with 
the lateral wall and top made of polycarbonate, and the 
base made of aluminum. The diameter of the box is 52 mm. 
Care was taken to avoid electrostatic effects by sprinkling 
the box with an antistatic spray and changing the beads
whenever they are seen to stick to the wall. The box 
is attached to the shaft of a shaker. The electromagnetic 
shaker (TiraVib 52110) is able to deliver a sinusoidal 
vibration to the box with a distortion smaller than 1\%, 
and its stiffness in the transversal direction is such 
that the residual acceleration is less than 0.05 $\mathbf{g}$ 
typically. The shaker is driven by an amplifier that is 
in turn commanded by a function generator, from where 
the frequency and amplitude are set. Two accelerometers 
are attached to the box to measure acceleration in the
vertical and one horizontal direction. The acceleration 
is acquired by a digital oscilloscope. Both the function 
generator and the oscilloscope are linked to a PC computer. 
An image acquisition system is also employed. It consists 
of a standard CCD camera with a macro lens. Images are 
recorded on a commercial VCR for subsequent treatment.

The procedure to measure the velocities involves the
acquisition of the bead traces under proper illumination.
A spatiotemporal diagram is obtained by registering a 
vertical line of pixels at regular time intervals and 
stacking them. The angle of the traces yields the bead
velocities. In order to measure the average angle we
perform a Fourier transform of the image and find the
position of the peaks. All errors combined yield to an
uncertainty of about 10\% at most.

We have observed that the speed of the beads near the
walls depends on height. They move faster in the upper
zone and their velocity decreases as they approach the
bottom, where the grains slow down and are entrained
into the bulk ---the velocity has a radial component to
close the flow. Therefore we have taken our measurements
in the upper third of the granular layer. We checked
that the bead velocity there does not depend on height, 
their radial velocity being negligible.
In that zone it is possible to track a bead and to obtain
a measurement of the downward velocity. Nevertheless, 
unavoidably the beads suffer small displacements in the 
azimuthal direction due to the rearrangements of the
layer, and so the traces have a small dispersion.
The source of error coming from the dispersion of 
the velocities adds to the fact that near the bottom 
the beads leap downward when another bead beneath is 
absorbed into the bulk. 

Several different granular materials have been used. The 
results we are reporting here have been obtained with glass 
beads of $0.5 \pm 0.1$ mm in diameter. Nevertheless, the 
same phenomena are also observed in sand, for example, with 
grains approximately that large but much more variable in 
size and form. Several layer depths were explored in the 
range $20<H<70$, where $H$ is the layer depth normalized by 
the bead diameter. In addition, the measurements shown here 
have been carried out at a frequency $f$=110 Hz, although in 
a wide range (from about 70 Hz to 120 Hz) the behavior was 
qualitatively similar. The non-dimensional acceleration 
$\Gamma=A\omega^2/\textbf{g}$ (where $A$ is the amplitude and 
$\omega=2\pi\, f$ the angular frequency) for which measurements were made ranges 
from about $0.5$ to $7$. For $\Gamma > 7$ the movement becomes 
rather disordered to make meaningful measurements.

Let us show the sequence of events for $H=20$, in the 
understanding that a similar picture emerges for other layer 
depths. When $\Gamma$ is small, the beads begin to move 
uniformly down the wall. 
Usually a small heap (much shorter than the height of the 
layer) is formed. If the box is well leveled 
(we do it to within $\pm$ 0.5 mrad) the heap appears at 
the center of the layer. As $\Gamma$ is increased, the heap 
decreases in height and by $\Gamma \ge 3 $ the layer
surface is almost flat; it just curves down near the
wall. At the same time, the velocity grows larger 
and larger. At this stage, the value of the velocity near 
the wall does not depend on the azimuthal position along 
the wall.

A further increase of acceleration produces a bifurcation 
towards another pattern: two rolls appear, breaking the 
circular symmetry of the previous planform (see Fig. 1). 
The acceleration 
threshold shows little ---if any--- dependence on the 
frequency in the range explored; it is about 
$\Gamma=6$. This transition, as remarked by Aoki and 
coworkers \cite{Aoki2}, presents hysteresis: once the rolls 
have formed, if acceleration is decreased they revert to 
the previous pattern at a lower value of $\Gamma$. 
The orientation of the rolls in the circular box from 
one run to another is seemingly random.

A profile of the velocities along the  azimuthal direction 
is shown in Fig. 2 before and after the bifurcation.
The spatial modulation of the velocity field is 
clearly noticeable. The projection of the velocities onto 
the vertical cross section of the cylinder is well fitted 
by a sine. The resemblance of these two patterns with 
those found in small aspect ratio B\'enard convection is 
striking. When a fluid layer open to the atmosphere 
heated from below begins to move, fluid goes up at the 
center and descends near the wall. If the temperature at 
the bottom is increased, a pair of rolls appear forming 
a pattern quite similar to the one found in granular 
materials \cite{Thierry}. The spatial patterns obtained 
bear a remarkable similarity. 
This strongly suggest that the convective motion in a 
shaken granular layer may be governed, at least in principle, 
by the symmetry restrictions imposed by the container. In 
order to provide the complete 
boundary conditions, the velocity field at the bottom of 
the container is needed. Work is in progress to obtain it 
following the same method.

In the range of frequencies and heights explored, we 
have always observed this transition from a circular 
pattern to two rolls. Further increase of the 
acceleration (explored only for some values of $H$) 
leads to an increase in the number of rolls, as 
described in previous experiments \cite{Aoki2}. 
The rolls are superimposed to the downward flow
at the wall (see Fig. 2 for $\Gamma=6.10$; note 
that the mean velocity is not zero).

In order to investigate the dependence of the velocity 
$\mathbf{v}$ near the wall  with the acceleration, we will 
focus in the 
first pattern, the one that appears for small $\Gamma$. 
This planform is concordant with the 
cross-section profiles obtained by Knight et al. 
\cite{Jaeger}. The velocity field takes the form of a 
torus, the grains rising at the center and going down 
near the walls. As it is not dependent of the azimuthal 
coordinate, a single value of the velocity determines 
the field. The dependence of $\mathbf{v}$ on the acceleration is 
shown in Fig. 3a for two layer depths: $H=20$ and $H=40$. 
The velocity is larger for thicker layers, and it grows 
monotonously with $\Gamma$. The relationship between $\mathbf{v}$
and $\Gamma$ near the threshold hints at an imperfect
transcritical bifurcation (the growth of $\mathbf{v}$ is almost
linear, as seen in Fig. 3b). The numerical results of
Ram\'{\i}rez et al., who propose a ``thermal convection''
mechanism, predict a supercritical bifurcation 
towards their first convective state. Nevertheless,
the first convective state they found is banned in
the experiment for symmetry reasons, and besides
their prediction is made for mean values of the
velocity.

It is noticeable that before the
transition from one to several rolls (for $\Gamma < 6$)
two different regimes exist with two distinct slopes
(Fig. 3 b). The changing of slope is more evident 
for the series corresponding to $H=20$. This behavior is 
more clearly displayed if the velocity $\mathbf{v}$ is 
normalized --dividing it by the amplitude of the velocity 
of the container wall $\mathbf{v_w}$. In doing this we 
are assuming that somehow the movement of the wall is 
driving the flow. The normalized velocity $\mathbf{v}
/ v_w$ is shown in Fig. 4 as a function of $\Gamma$. 
Clearly all series now give the same behavior: 
a first stage where $\mathbf{v}/v_w$ grows with $\Gamma$ 
and a second one where it remains almost constant. This 
saturation strongly suggests an effect related to 
friction, a force that depends on the velocity. The
friction could either be internal (among the grains
themselves) or external (between grains and wall). It
is difficult to make an educated guess to decide which
is the case; additional work is needed to further
clarify this point. The mechanisms invoked by Knight
et al. \cite{Nagel} and by Ram\'{\i}rez et al.
\cite{Ramirez} are both compatible in principle with
such a fact.

An additional rescaling can be done dividing the 
normalized velocity by $H$. After doing this, all the 
curves collapse in a single one (Fig. 5). This 
dependency with $H$ is revealing, but again it is
difficult to decide whether it is the friction with
walls or the ``granular temperature'' the driving
force. The first one should yield a scaling with $H$ 
for a fixed cross section, as it is the case. The 
second one could also give a scaling with $H$, although
the reason why it should scale \emph{linearly} is 
---to our knowledge--- unexplained.
 
In summary, we have presented the results of an 
experiment in which the velocity of the grains near 
the walls has been measured locally. This allows 
a quantitative description of the convecting 
patterns. Moreover, the behavior of the velocity shows 
two different behaviors as the acceleration $\Gamma$ 
is increased; this strongly suggests that a mechanism
associated with friction is involved. The scaling of
the velocity with $\mathbf{v_w}$ and $H$ favors the case of a
driving force associated with wall friction, although
``thermal convection'' cannot be excluded.

This work has been partially funded by Project PIUNA
from the Universidad de Navarra, Project PB98-0208 
from DGYCIT (Spanish Ministerio de Educaci\'on) and 
Project HPRN-CT-2000-00158 from the European Community. 
We thank S. Boccaletti, W. Gonz\'alez, H. Mancini 
and D. Valladares for their useful comments.
JLI acknowledges AECI for financial support, and
IZ thanks Asociaci\'on de Amigos de la Universidad
de Navarra for a grant.

\newpage
\vspace{-0.5cm}
\begin{small}

\end{small}

\pagebreak

\begin{center}

{\bf Figure Captions}

\begin{itemize}

\item[]{\bf Figure 1} The surface of the convecting
granular material under vertical vibration ($\Gamma=6.10$,
$f=110 Hz$, $H=20$).
The lateral illumination reveals the prominences of the
relief, showing a planform of two rolls that protrude
slightly at their axes. 
The surface of the pattern consisting of a single toroidal
roll is almost featureless, except for a small heap at
small accelerations.

\item[]{\bf Figure 2} Velocity profiles near the wall
as a function of the azimuthal angle $\theta$:
circles, $\Gamma=2.01$; triangles, $\Gamma=3.07$; 
squares, $\Gamma=6.10$. (This last series is noisy 
but it has the virtue that by chance the maximum 
and minimum of $\mathbf{v}$ are seen). Note that only a 
portion of the complete circumference is accessible 
(about 160$^{\circ}$). All the velocities are 
downward. It can be seen that $\mathbf{v}$ is constant for 
small $\Gamma$ --the convective pattern is a single 
toroidal roll-- while for larger $\Gamma$ an spatial
oscillation is revealed --two rolls are formed, as
shown in Fig. 1.
The measurements were taken at $f=110 Hz$ and $H=20$. 

\item[]{\bf Figure 3}
The dependence of $\mathbf{v}$ on $\Gamma$ hints to an imperfect
transcritical bifurcation (a), where a linear growth is
clearly apparent for small acceleration (b). For a larger 
range of accelerations (but still corresponding only to
the first pattern, i.e. the toroidal convective roll, for
which $\mathbf{v}$ does not depend on the azimuthal coordinate)
 two linear regimes are seen.
In the second one, the slope is smaller. Squares 
correspond to $H=40$ and triangles to $H=20$. The data
shown correspond to several runs, some of
them increasing $\Gamma$ and others decreasing it: no
hysteresis is found in $\mathbf{v}$. Measurements were made at
$f=110 Hz$. The fits are just guides for the eye.

\item[]{\bf Figure 4}
The normalized velocity $\mathbf{v / v_w}$, where $\mathbf{v_w}$ is the 
maximum velocity of the wall, as a function of $\Gamma$. 
Squares: $H=20$; circles: $H=80/3$; down triangles: $H=100/3$;
up triangles: $H=30$; diamonds: $H=140/3$. The fits (logistic
functions) are guides for the eye.

\item[]{\bf Figure 5}
The rescaled velocity $\mathbf{\frac{v}{v_w H}}$ as a function
of $\Gamma$. Data are the same than in Fig. 4, all 
combined. The logistic fit is a guide for the eye.

\end{itemize}

\end{center}

\begin{thebibliography}{}

\bibitem{Jaeger}
H.M. Jaeger and S.R. Nagel, 
Science {\bf 255}, 1523 (1992).
\bibitem{Duran}
J. Duran, {\it Sables, poudres et grains}, 
Ed. Eyrolles, Paris (1997).
\bibitem{Nagel}
J.B. Knight, E.E. Ehrichs, V.Y. Kuperman, J.K. Flint, 
H.M. Jaeger and S.R. Nagel,
Phys. Rev. E {\bf 54}, 5726 (1996), and references therein.
\bibitem{Cross}
M.C.Cross and P.C. Hohenberg, 
Rev. Mod. Phys. {\bf 65} 1361 (1993).
\bibitem{Melo}
F. Melo, P. Umbanhowar and H.L. Swinney, 
Phys. Rev. Lett. {\bf 72}, 172 (1994).
\bibitem{Laroche}
C. Laroche, S. Douady and S. Fauve, 
J. Phys. France {\bf 50}, 699 (1989);
S. Fauve, S. Douady and C. Laroche, 
J. Phys. France {\bf 50}, C3-187 (1989).
\bibitem{Aoki2}
K.M. Aoki, T. Akiyama, Y. Maki and T. Watanabe,
Phys. Rev. E {\bf 54}, 874 (1996).
\bibitem{Rajchenbach1}
J. Rajchenbach, Europhys. Lett. {\bf 16}, 149 (1991);
see also E. Cl\'ement and J. Rajchenbach, 
Europhys. Lett. {\bf 16}, 133 (1991).
\bibitem{Ramirez}
Rosa Ram\'{\i}rez, D. Risso, and P. Cordero ,
Phys. Rev. Lett. {\bf 85}, 1230 (2000) 
\bibitem{Behringer}
H.K. Pak, E. Van Doorn and R.P. Behringer,
Phys. Rev. Lett. {\bf 74}, 4643 (1995).
\bibitem{Thierry}
T. Ondar\c{c}uhu, J. Mill\'an-Rodr\'{\i}guez, H.L. Mancini, 
A. Garcimart\'{\i}n and C. P\'erez-Garc\'{\i}a, 
Phys. Rev. E {\bf 48}, 1051 (1993). M.L. Ram\'on, 
D. Maza and H.L. Mancini, Phys. Rev. E {\bf 60}, 4193 (1999). 

\end{thebibliography}
\end{document}